\documentclass[a4paper,fleqn,usenatbib]{mnras}

\usepackage[T1]{fontenc}
\usepackage{ae,aecompl}

\usepackage{graphicx}   
\usepackage{amsmath}    
\usepackage{amssymb}    

\title[East-West Longitudinal Asymmetry of SEPs]{On the East-West Longitudinally Asymmetric Distribution of Solar Proton Events}

\author[H.-Q. He and W. Wan]{H.-Q. He$^{1}$\thanks{E-mail: hqhe@mail.iggcas.ac.cn} and W. Wan $^{1}$ \\
$^{1}$ Key Laboratory of Earth and Planetary Physics, Institute of
Geology and Geophysics, Chinese Academy of Sciences, Beijing 100029,
China}

\pubyear{2016}

\begin{document}
\label{firstpage}
\pagerange{\pageref{firstpage}--\pageref{lastpage}}
\maketitle

\begin{abstract}
A large data set of 78 solar proton events observed near the Earth's
orbit during 1996-2011 is investigated. An East-West longitudinal
(azimuthal) asymmetry is found to exist in the distribution of flare
sources of solar proton events. With the same longitudinal
separation between the flare sources and the magnetic field line
footpoint of observer, the number of the solar proton events
originating from solar sources located on the eastern side of the
nominal magnetic footpoint of observer is larger than the number of
the solar proton events from solar sources located on the western
side. We emphasize the importance of this statistical investigation
in two aspects. On the one hand, this statistical finding confirms
our previous simulation results obtained by numerically solving
five-dimensional Fokker-Planck equation of solar energetic particle
(SEP) transport. On the other hand, the East-West longitudinally
(azimuthally) asymmetric distribution of solar proton events
accumulated over a long time period provides an observational
evidence for the effects of perpendicular diffusion on the SEP
propagation in the heliosphere. We further point out that, in the
sense of perpendicular diffusion, our numerical simulations and
statistical results of SEP events confirm each other. We discuss in
detail the important effects of perpendicular diffusion on the
formation of the East-West azimuthal (longitudinal) asymmetry of SEP
distribution in two physical scenarios, i.e., ``multiple SEP events
with one spacecraft" and ``one SEP event with multiple spacecraft".
A functional relation $I_{max}(r)=kr^{-1.7}$ quantifying the radial
dependence of SEP peak intensities is obtained and utilized in the
analysis of physical mechanism. The relationship between our results
and those of Dresing et al. is also discussed.
\end{abstract}




\section{Introduction}

Solar energetic particles (SEPs) are charged energetic particles
occasionally emitted by the Sun during its burst events. Generally,
SEPs are produced by solar flares or/and related to coronal mass
ejections (CMEs), although the relative importance of flares and
CME-driven shocks in producing high-energy particles is not
completely understood. The propagation of SEPs in the heliosphere
remains one of the important subjects of space physics,
heliophysics, and plasma physics. In addition, large solar proton
events (SPEs) significantly affect the solar-terrestrial space
environment, and thereby have become an important topic in the
research fields of space radiation, space weather, and space
climate. Recently, multi-spacecraft observations (STEREO-A/B, ACE,
SOHO, Wind, etc.) and three-dimensional transport modeling have been
intensely used in the studies of SEP events.

Essentially, the propagation of SEPs in interplanetary space
includes several fundamental physical mechanisms, such as particle
streaming along magnetic field lines, convection with the solar
wind, adiabatic magnetic focusing, pitch-angle diffusion, adiabatic
cooling, and perpendicular diffusion. According to the well-known
quasi-linear theory for cosmic ray diffusion \citep{Jokipii1966},
the perpendicular diffusion coefficient is usually much smaller than
the parallel diffusion coefficient. Therefore, the effects of
perpendicular diffusion were often ignored in previous theoretical,
numerical, and observational studies of SEP propagation. However,
the observations of SEP events detected by the Helios mission
proposed some nonlinear effects near $90^{\circ}$ pitch angle of
particles
\citep[e.g.,][]{Hasselmann1968,Hasselmann1970,Beeck1986,Beeck1987}.
Further investigations showed that when the magnetic turbulence is
quite strong, the perpendicular diffusion could be comparable to the
parallel diffusion \citep[e.g.,][]{Shalchi2005}. Recently, with the
launch of the twin spacecraft STEREO-A/B and the development of
three-dimensional transport modeling, the significant role of
perpendicular diffusion in the SEP propagation has become gradually
and increasingly well known in the SEP community. The important
effects of perpendicular diffusion on SEP transport and distribution
in the heliosphere were found in both numerical modeling studies and
observational studies
\citep[e.g.,][]{Zhang2009,He2011,He2015a,He2015b,Droge2010,Droge2014,Dresing2012,Dresing2014,Giacalone2012,Strauss2015}.
Note that the diffusion could not only be anisotropic with respect
to the directions perpendicular and parallel to the magnetic field,
but that it could also be fully anisotropic. For example, the case
of two different perpendicular diffusion coefficients has been
discussed in \citet{Effenberger2012}, and also applied to SEPs by
\citet{Kelly2012}. In the sense of perpendicular diffusion, some
interesting SEP phenomena were reproduced or found by performing
complete model calculations of SEP propagation in the
three-dimensional interplanetary magnetic field with the effects of
turbulence \citep[for a recent brief review, see][]{He2015b}.

In previous simulation works, it was found that with the same
heliographic longitude separation between the magnetic footpoint of
the observer and the solar sources, the SEPs associated with the
sources located at east are detected earlier and with larger fluxes
than those associated with the sources located at west
\citep{He2011,He2015a}. This SEP phenomenon was called the
``East-West azimuthal asymmetry" by \citet{He2011}. In this work, we
focus on the so-called East-West azimuthal asymmetry of SEPs in view
of statistical analysis of solar proton events. We mainly pay
attention to the relatively longitudinal distribution of solar
proton events. The longitudinally asymmetric distribution phenomenon
of solar proton events is found. With the same longitudinal
separation between magnetic footpoint of observer and solar sources,
the number of the SPEs originating from sources located on the
eastern side of the magnetic footpoint of observer is larger than
the number of the SPEs from sources located on the western side.
Essentially, this observational result is consistent with our
previous numerical simulation results. We further discuss in detail
the physical mechanism responsible for this SEP phenomenon in two
scenarios, i.e., ``multiple SEP events with one spacecraft" and
``one SEP event with multiple spacecraft". The effects of
perpendicular diffusion on the heliospheric propagation and the
East-West longitudinally (azimuthally) asymmetric distribution of
SEPs are emphasized in the discussions. We also compare our
statistical and numerical results with the observational study of
\citet{Dresing2014}. In the context of perpendicular diffusion, our
results and those of \citet{Dresing2014} are unified and identified
with each other.

This paper is structured as follows. In Section 2, we briefly
describe the collected data set of 78 solar proton events
accumulated during 1996-2011, based on which the statistical
analysis will be carried out. In Section 3, we present the
statistical results of the 78 solar proton events. In Section 4, we
discuss the physical mechanism responsible for the East-West
longitudinal (azimuthal) asymmetry of SEP distribution in the
heliosphere. The important effects of perpendicular diffusion will
be discussed in detail. The inherent relationship between our works
and other statistical survey of SEP events \citep{Dresing2014} will
be discussed in Section 5. A summary of our results will be provided
in Section 6.

\section{Solar Proton Events Analyzed in This Investigation}

The 78 solar proton events affecting the Earth environment detected
during 1996-2011 are collected from Space Weather Prediction Center
of National Oceanic and Atmospheric Administration (NOAA). The solar
proton events and associated flares and CMEs are listed in Table
\ref{table:events}. The maximum intensity of each solar proton event
for energies $>10$ MeV is larger than 10 particle flux units (pfu).
Note that 1 pfu=1 particles/($cm^{2}-sr-s$). We also note that the
78 solar proton events presented in Table \ref{table:events} are
collected from the original listing by excluding the events without
identifiable origin location on the solar surface. The SEP events
without definite source location are randomly distributed on the
solar surface, i.e., the SEP events excluded from the listing may
locate on either the eastern or the western side of the solar
surface.

The first three columns in Table \ref{table:events} give the start
time, maximum time, and proton flux of each SEP event, respectively.
The main direction and onset time of CME associated with each
particle event are listed in Column 4. Columns 5 and 6 provide the
maximum time and importance (X ray/optical) of flare associated with
each SEP event. The majority of the SEP events presented in Table 1
were observed at the onset of the solar events. Generally, intense
SEP events are usually associated with both major flares and large
CMEs, and consequently, the relative roles of flares and CME-driven
shocks in producing high-energy particles are not completely
understood \citep{Cliver2002}. In addition, most of the SEPs in
impulsive events are released near the surface of the Sun. Even in
many gradual events, energetic particles, particularly those with
high energies, are generated near the Sun, where the shock is very
fast, the magnetic field is quite strong and the seed particles are
dense \citep{Zhang2009}. Therefore, in the investigation, we do not
particularly discriminate between the two types of particle sources.
The location and number of each associated active region are listed
in columns 7 and 8, respectively. The last three columns in Table
\ref{table:events} provide the solar wind speeds measured by
spacecraft Solar and Heliospheric Observatory (SOHO), ACE, and the
mean value of them, respectively. By averaging SOHO or ACE data
measured in the time interval $[T_{max}-2 hr, T_{max}+2 hr]$, in
which $T_{max}$ indicates the maximum time of the corresponding SEP
event, we can obtain the solar wind speed in each SEP event. The
averaged solar wind speed will be used in our statistical
investigation. As one can see, for some solar proton events, only
the SOHO data of solar wind speeds are available, and for some
events, only the ACE data are available. Moreover, if there are not
valid solar wind speed data in the time range $[T_{max}-2 hr,
T_{max}+2 hr]$ for a certain solar proton event, the averaged solar
wind speed obtained from the complete observational data on the
whole day when the maximum of the event occurred will be used. In
addition, we note that the solar wind speed $790~km~s^{-1}$ during
the 2003 October 28 solar proton event listed in Table
\ref{table:events} is adopted from \citet{Veselovsky2004}.

In the investigation, we utilize the solar wind speeds according to
the rules as: if both of the solar wind speeds obtained from SOHO
and ACE data, respectively, are available, then we employ the mean
value of them, i.e.,
$\overline{V}^{sw}=(V^{sw}_{soho}+V^{sw}_{ace})/2$, where
$V^{sw}_{soho}$ and $V^{sw}_{ace}$ are the solar wind speeds
averaged from SOHO and ACE data, respectively; otherwise, the solar
wind speed derived from the measurements by a single spacecraft
(SOHO or ACE) will be used.

\section{East-West Azimuthal Asymmetry of Solar Proton Events}

The rotation angle of the nominal interplanetary magnetic field line
connecting the observer at radial distance $r$ with the Sun can be
calculated through this expression \citep[e.g.,][]{Parker1958}
\begin{equation}
\phi_{s}=\Omega r/V^{sw}. \label{spiral-angle}
\end{equation}
Here, $\Omega$ indicates the Sun's angular rotation rate, $r$
indicates the heliocentric radial distance, and $V^{sw}$ indicates
the solar wind speed. Note that the nominal magnetic footpoint of
the spacecraft calculated through the above expression may differ
from the actual footpoint, since sometimes the deviations of the
interplanetary magnetic field can be significant. But in the
statistical sense, this will not affect our main results. The west
and east heliographic longitudes $\phi$ of the solar sources are
usually indicated by positive and negative values, respectively. The
relative longitude $\phi_{r}$ between the solar flare and the
observer's magnetic footpoint can be obtained via this relation
\begin{equation}
\phi_{r}=\phi-\Omega r/V^{sw}. \label{relative-longitude}
\end{equation}
According to the realistic solar wind speed measured by the
spacecraft, we can obtain the relative longitude $\phi_{r}$ for each
solar proton event listed in Table \ref{table:events}.

\begin{figure}
    \centering
    \includegraphics[width=\columnwidth]{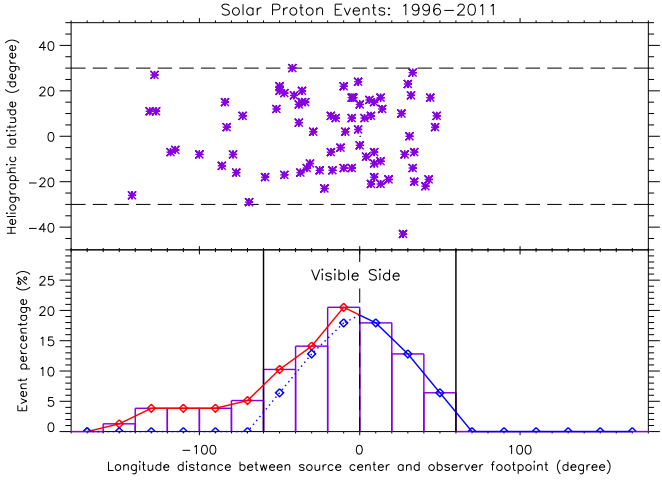}
    \caption{Spatial distribution of the solar sources of 78 solar proton events
    observed near the Earth's orbit during 1996-2011. Upper panel: latitudinal and longitudinal distribution. Lower panel: number
percentage distribution of the solar events along the relative
longitude $\phi_{r}$. Adapted from Figure 1 of \citet{He2013}.}
    \label{observation}
\end{figure}

We present the statistical result of the solar proton events in
Figure \ref{observation}. A preliminary result was reported in
\citet{He2013}. In the upper panel, it can be seen that the solar
flares are primarily distributed within $[S30^{\circ},
N30^{\circ}]$, which is consistent with the result in
\citet{Zhao2007}. We divide the relative longitude $\phi_{r}$ into
bins: $[-180^{\circ}, -160^{\circ}]$, $[-160^{\circ},
-140^{\circ}]$, $\ldots$, $[160^{\circ}, 180^{\circ}]$. The number
percentage distribution of the solar events along the relative
longitude $\phi_{r}$ is presented in the lower panel. We note that
the negative value (left) and positive value (right) indicate that
the solar flares locate on the eastern side and western side of the
magnetic footpoint of observer, respectively. To directly compare
the event number percentages between the east and west relative
longitude, the event number percentages on the right side are
mirrored to the left side, indicated by a blue dotted line. We
mainly pay attention to the central longitude range of the visible
side of the Sun, as indicated between the two vertical lines in
Figure \ref{observation}. It can be seen that with the same
longitudinal separation between the magnetic footpoint of observer
and solar sources, the number of the solar proton events originating
from sources located on the eastern side of the observer's magnetic
footpoint is larger than the number of the solar proton events from
sources located on the western side \citep{He2013}.

\begin{figure}
 \centering
 \includegraphics[width=\columnwidth]{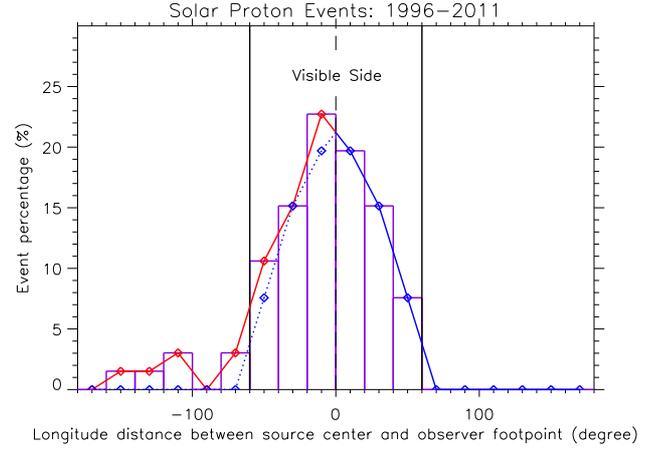}
 \caption{Number percentage distribution of the solar sources of the 66 SEP events
with relatively short delays along the relative longitude
$\phi_{r}$. With the same longitudinal distance between the
observer's magnetic footpoint and solar sources, the number of the
SEP events originating from sources located on the eastern (left)
side of the observer's footpoint is larger than the number of the
SEP events from sources located on the western (right) side.}
      \label{observation-66}
\end{figure}

Generally, a SEP event which starts with a long delay at the
observer is more possibly related to a CME-driven shock. So we
further collect 66 SEP events from Table \ref{table:events} by
excluding the events with a delay of more than 24 hours with respect
to the flare maximum. We also statistically investigate the number
percentage distribution of these 66 SEP events along the relative
longitude $\phi_{r}$. Figure \ref{observation-66} shows the
statistical results. We also pay attention to the central longitude
range of the visible side of the Sun. As we can see, the number
percentage of the SEP events on the left side is generally larger
than that on the right side, similar to the results in Figure
\ref{observation}. As noted above, most of the high-energy SEPs
(associated with either flares or CME-driven shocks) are probably
produced near the Sun. After the SEPs are released on or near the
Sun, they will transport in the interplanetary magnetic field with
turbulence. During the transport processes in the heliosphere, the
parallel diffusion and perpendicular diffusion play a very important
role in the SEP distribution and anisotropy. Therefore, we note that
even for CME-associated SEP events, the East-West longitudinal
(azimuthal) asymmetry of SEP distribution, due to the effects of
perpendicular diffusion, will hold. In this sense, it is not very
necessary to particularly discriminate between the two types of SEP
sources, i.e., solar flares and CME-driven shocks. Actually, the
traditional classification paradigm of the so-called impulsive and
gradual SEP events is increasingly challenged by the current
multi-spacecraft observations (STEREO-A/B, ACE, SOHO, Wind, etc.)
and the numerical modeling of three-dimensional focused transport of
SEPs.

\section{Discussions on the Mechanism Responsible for East-West Azimuthal Asymmetry of SEPs}

\begin{figure}
 \centering
 \includegraphics[width=\columnwidth]{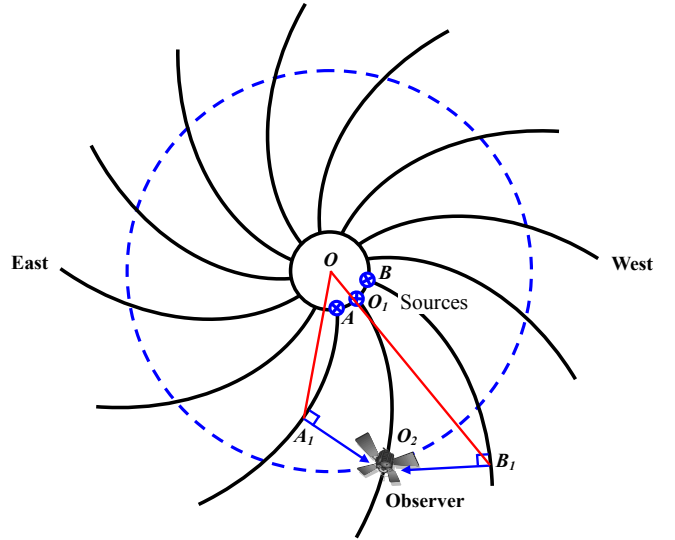}
 \caption{Diagram to illustrate the formation of the East-West
 longitudinal (azimuthal) asymmetry of SEP distribution via perpendicular
 diffusion in the physical scenario of ``multiple SEP events with one spacecraft".
 In a statistical sense, the intensity of SEPs (detected at position $O_{2}$)
 originating from solar source A (B) located on the eastern (western) side of the
 observer footpoint is nearly equivalent to the SEP intensity at position $A_{1}$ ($B_{1}$).
 The heliocentric radial distance $\overline{OA_{1}}$ between solar center (O) and the position
$A_{1}$ is shorter than that $\overline{OB_{1}}$ between solar
center (O) and the position $B_{1}$.}
      \label{Illustration-He}
\end{figure}

Our previous numerical simulations of SEP transport with the effect
of perpendicular diffusion demonstrated that there exists an
East-West longitudinal (azimuthal) asymmetry both in the entire
intensities and in the peak intensities of SEPs transporting in the
interplanetary space \citep[see][]{He2011,He2013,He2015a}. As
pointed out by these works, the longitudinally (azimuthally)
asymmetric distribution of SEPs results from the East-West azimuthal
asymmetry in the geometry of the Parker interplanetary magnetic
field as well as the effects of perpendicular diffusion on the
transport processes of SEPs in the heliosphere. The formation
mechanism of the East-West longitudinal (azimuthal) asymmetry of SEP
distribution via perpendicular diffusion is sketched in Figure
\ref{Illustration-He}, which is referred to the physical scenario of
``multiple SEP events with one spacecraft". The SEP transport in the
interplanetary magnetic field mainly includes the parallel diffusion
and the perpendicular diffusion. Due to the effect of the
perpendicular diffusion, the SEPs released from acceleration regions
with limited coverage can cross the interplanetary magnetic field
lines and transport to distant heliospheric locations with very wide
longitudinal or/and latitudinal separations. In Figure
\ref{Illustration-He}, the intensity of SEPs (detected at
spacecraft's position $O_{2}$) originating from solar source A (as
an example) located on the eastern side of the observer footpoint is
nearly equivalent to the SEP intensity at position $A_{1}$, in a
statistical sense. Similarly, the intensity of SEPs (detected at
spacecraft's position $O_{2}$) released from solar source B (with
the same longitudinal separation as source A relative to the
observer footpoint) located on the western side is nearly equivalent
to the SEP intensity at position $B_{1}$, in a statistical sense. As
one can see, the heliocentric radial distance $\overline{OA_{1}}$
between solar center (O) and the position $A_{1}$ is shorter than
that $\overline{OB_{1}}$ between solar center (O) and the position
$B_{1}$. Note that the extent of the length difference between the
heliocentric radial distances $\overline{OA_{1}}$ and
$\overline{OB_{1}}$ depends on the longitudinal separation between
the sources and the observer's footpoint. Specifically, a larger
longitudinal separation leads to a more significant length
difference between the radial distances of the east and west
equivalent positions (e.g., $A_{1}$ and $B_{1}$ in Figure
\ref{Illustration-He}).

\begin{figure}
 \centering
 \includegraphics[width=\columnwidth]{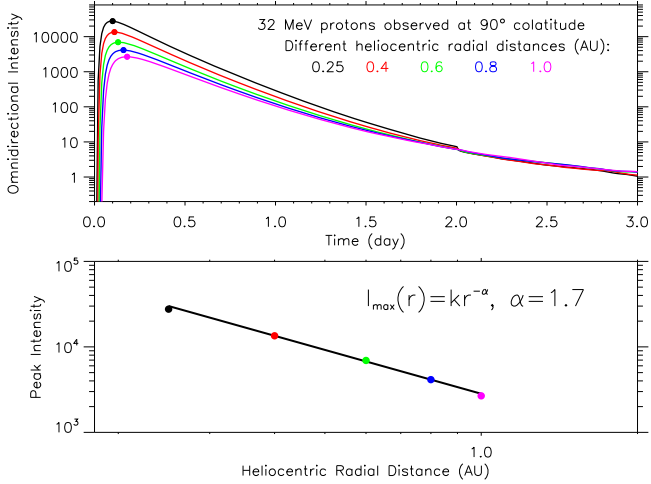}
 \caption{Simulated radial dependence of intensity-time profiles and peak
intensities of SEPs observed along the nominal Parker magnetic field
line connecting the solar source. Upper panel: The intensity-time
profiles of $32$ MeV protons observed at different radial distances:
$0.25$, $0.4$, $0.6$, $0.8$, and $1.0$ AU. Both the solar source and
the observers are located at $90^{\circ}$ colatitude. Note that the
so-called SEP reservoir phenomenon is reproduced in the simulations
of a series of SEP cases with different radial distances. Lower
panel: The radial dependence of the SEP peak intensities extracted
from the intensity-time profiles shown in the upper panel. A
functional form of $I_{max}(r)=kr^{-\alpha}$ with index $\alpha=1.7$
is obtained.}
      \label{Imax-radial}
\end{figure}

Observational analysis showed that the peak intensities of SEPs
decrease with the heliocentric radial distance (r) in a functional
form of $r^{-\alpha}$, with $\alpha$ varying in a wide range.
Basically, our numerical modeling of the five-dimensional
Fokker-Planck transport equation has confirmed these observational
results. In this paper, we present a numerical modeling result in
Figure \ref{Imax-radial} regarding radial dependence of
intensity-time profiles and peak intensities of SEPs observed along
the nominal Parker magnetic field line connecting the spacecraft
with the solar source. The upper panel of Figure \ref{Imax-radial}
presents the intensity-time profiles of $32$ MeV protons observed at
different heliocentric radial distances: $0.25$, $0.4$, $0.6$,
$0.8$, and $1.0$ AU. Both the solar source and the observers are
located at $90^{\circ}$ colatitude in this typical case study. As
one can see, during the entire evolution process, the particle
intensity observed at a smaller radial distance is higher than the
particle intensity detected at a larger radial distance.

In addition, we can obviously see that during the late phase, the
intensity-time profiles of all the SEP cases display nearly the same
decay rate, which is known as the SEP reservoir phenomenon in
spacecraft observations \citep[e.g.,][]{Roelof1992}. Therefore, our
simulations of a series of SEP cases with different radial distances
reproduce the so-called SEP reservoir phenomenon. Actually, our
numerical simulations with perpendicular diffusion reproduced
various SEP reservoirs at different heliospheric locations
(longitude, latitude, and radial distance) without invoking the
prevailing hypothesis of a reflecting boundary (magnetic mirror) or
diffusion barrier (solar wind structure)
\citep[e.g.,][]{He2011,He2015a,He2015b}. Additionally, so far no any
explicit observational evidence or quantitative description of
reflecting boundaries (magnetic mirrors) or diffusion barriers
(solar wind structures), which were once prevalently supposed to
exist in the heliosphere to contain the particles long enough to
form the observed SEP reservoirs, has been provided or specified
within the community \citep[for discussions, see][]{He2015b}.
Actually, as \citet{He2015a} and \citet{He2015b} pointed out, it is
difficult to imagine such an ``overwhelming" reflecting boundary
(magnetic mirror) or diffusion barrier (solar wind structure)
covering all of the longitudes, all of the latitudes, and even all
of the radial distances, and even exactly accompanying almost all of
the SEP events including the so-called impulsive or $^{3}$He-rich
events. Therefore, in the sense of perpendicular diffusion, we
suggest that the so-called SEP ``reservoir" should be renamed SEP
``flood", which is more appropriate to illustrate the transport
processes of SEPs in the heliosphere.

In the upper panel of Figure \ref{Imax-radial}, the filled circles
with different colors on the intensity-time profiles denote the peak
intensities of the corresponding SEP cases. We extract the
information of the peak intensities and the relevant radial
distances in the SEP cases, and present it in the lower panel of
Figure \ref{Imax-radial}. We further model the radial dependence of
the peak intensities $I_{max}$ with a power-law function
$I_{max}(r)=kr^{-\alpha}$, where $k$ is a constant in this case
study. We obtain the power-law index $\alpha=1.7$ in this series of
SEP cases. We note that the index $\alpha$ mainly depends on the
properties (e.g., coverage, location) of SEP sources and the
relative locations of magnetic field line footpoints of the
observers. From the lower panel of Figure \ref{Imax-radial}, we can
clearly see that the peak particle intensities of SEP events
exponentially decrease with the increasing radial distances.

According to the relationship $\overline{OA_{1}}<\overline{OB_{1}}$
sketched in Figure \ref{Illustration-He} and the radial dependence
of SEP peak intensities $I_{max}(r)=kr^{-\alpha}$ with, e.g.,
$\alpha=1.7$ in this work, we can readily deduce that the peak
intensity observed at position $A_{1}$ is larger than that observed
at position $B_{1}$. Actually as shown in the upper panel of Figure
\ref{Imax-radial}, the SEP intensity detected at position $A_{1}$ is
larger than that detected at position $B_{1}$ through the whole
evolution process. We also note that these results are valid for
arbitrary energies of energetic particles and for arbitrary particle
species including electrons and heavy ions. Therefore, the particle
fluence and radiation dose received at position $A_{1}$ are,
respectively, higher than those received at position $B_{1}$. Then
according to the intensity equivalence relationship discussed above
between $A_{1}$ and $O_{2}$ (observing SEPs from source A) as well
as that between $B_{1}$ and $O_{2}$ (observing SEPs from source B),
we can infer that the entire intensity, peak intensity, particle
fluence, and radiation dose originating from source A are,
respectively, larger than those originating from source B. We note
that these conclusions hold for arbitrary couple of solar sources
located on the eastern and western sides of the observer's magnetic
footpoint, respectively, but with the same longitudinal separation
relative to the observer's magnetic footpoint.

In typical cases of interplanetary conditions, the perpendicular
diffusion coefficient should be smaller than the parallel diffusion
coefficient, as performed in our previous numerical modeling,
although the effect of perpendicular diffusion always plays a very
important role in the transport and distribution of SEPs. Therefore,
we note that the longitudinal separation between the source region
and the magnetic footpoint of the observer is the most crucial
factor in determining the longitudinal variation of the entire
intensity and the peak intensity of SEPs, indicating that the
farther the magnetic footpoint of the observer is away from the
solar source, the smaller the SEP intensity will be observed by the
spacecraft and also the later the SEP event onset will be detected.

The probability of an SEP event being detected by spacecraft in the
interplanetary space is related to the SEP intensity in the event
and the detection threshold of the instruments onboard the
spacecraft. In addition, note that the criterion of an event being
recorded in the SPE listing of NOAA is that the flux of the event
being larger than or equal to 10 pfu. Therefore, the spacecraft
detection threshold and SPE listing criterion naturally play a role
in the historical records of SEP events affecting the Earth
environment. Our previous simulations showed that with the same
longitudinal separations between the solar sources and the magnetic
footpoint of the observer, the SEP events originating from solar
sources on the eastern side of the magnetic footpoint of the
observer reveal larger fluxes than those from solar sources on the
western side \citep{He2011,He2015a}. Therefore, the SEP events from
solar sources on the eastern (western) side of observer's magnetic
footpoint will have a larger (smaller) probability of being detected
and recorded by spacecraft near the Earth's orbit. When a large data
set of SEP events is taken into account, it is natural and
expectable to obtain the East-West longitudinal (azimuthal)
asymmetry (Figures \ref{observation} and \ref{observation-66}). As
discussed above, the detection threshold and listing criterion
naturally play a role in the statistical results of the East-West
longitudinally (azimuthally) asymmetric distribution of SEP events.
We point out that no matter what the detection threshold (e.g.,
particle flux, particle energy, particle species, etc.) or listing
criterion is, the result of the East-West longitudinal (azimuthal)
asymmetry of SEP events will hold, provided that a relatively large
data set is taken into account. We emphasize that our numerical
simulations of five-dimensional transport equation and statistical
analysis of SEP events confirm each other. Furthermore, the
combination of statistical survey and numerical modeling regarding
the East-West longitudinal (azimuthal) asymmetry presents a unique
evidence for the effects of perpendicular diffusion on the SEP
propagation in the heliosphere.

\section{Relationship between Our Results and Those of a Recent Work}

In a recent work \citep{Dresing2014}, the authors also presented a
statistical survey of the longitudinal distribution of SEP peak
intensities observed by twin spacecraft STEREO-A/B and near-Earth
spacecraft ACE. A similar investigation of SEP peak intensities can
be found in \citet{Lario2013}, where the SEP intensity asymmetry was
also presented. Essentially, their results, obtained in a ``one SEP
event with multiple spacecraft" style, are consistent with and
confirm our numerical modeling and observational results. In this
section, we discuss the relationship between our results and those
of \citet{Dresing2014}.

\begin{figure}
 \centering
 \includegraphics[width=\columnwidth]{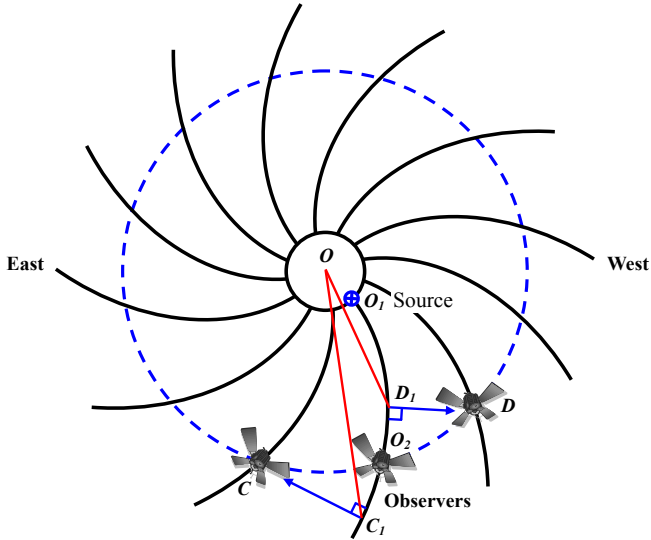}
 \caption{Diagram to illustrate the formation of the East-West longitudinal
(azimuthal) asymmetry of SEP distribution via perpendicular
diffusion in the physical scenario of ``one SEP event with multiple
spacecraft". In a statistical sense, the SEP intensity observed by
spacecraft D (C) located on the western (eastern) side is nearly
equivalent to the SEP intensity observed at position $D_{1}$
($C_{1}$). The heliocentric radial distance $\overline{OD_{1}}$
between solar center (O) and the position $D_{1}$ is smaller than
that $\overline{OC_{1}}$ between solar center (O) and the position
$C_{1}$.}
      \label{Illustration-Dresing}
\end{figure}

Figure \ref{Illustration-Dresing} sketches the physical scenario of
the multi-spacecraft observations of SEPs originating from a common
particle source (e.g., source $O_{1}$) on the solar surface. The
three spacecraft (marked with C, $O_{2}$, and D, respectively, from
East to West) are located at heliospheric positions with the same
heliocentric radial distance but with different heliolongitudes.
Specifically, the spacecraft $O_{2}$ in the middle is directly
connected to the solar source $O_{1}$ by the interplanetary magnetic
field line, whereas the magnetic footpoints of the spacecraft C on
the left and spacecraft D on the right are located on the eastern
side and western side, respectively, with the same longitudinal
separation, relative to the solar source $O_{1}$. After being
released from the solar surface, the energetic particles will
transport diffusively in turbulent interplanetary magnetic field,
consisting of parallel diffusion along and perpendicular diffusion
across the mean magnetic field. Due to the effects of perpendicular
diffusion, the energetic particles released from a solar source
(e.g., source $O_{1}$) with limited coverage can be detected by
multiple spacecraft with wide longitudinal or/and latitudinal
separations between every couple of them. The SEP intensity observed
by spacecraft D located on the western side is nearly equivalent to
the SEP intensity observed at position $D_{1}$, in a statistical
sense. Likewise, the SEP intensity detected by spacecraft C located
on the eastern side is nearly equivalent to the SEP intensity
detected at position $C_{1}$, in a statistical sense. As we can see,
the heliocentric radial distance $\overline{OD_{1}}$ between solar
center (O) and the position $D_{1}$ is smaller than that
$\overline{OC_{1}}$ between solar center (O) and the position
$C_{1}$. Note that the extent of the length difference between the
radial distances $\overline{OC_{1}}$ and $\overline{OD_{1}}$ depends
on the longitudinal separation between the source and the observer
footpoints. A larger longitudinal separation leads to a more
considerable length difference between the radial distances
$\overline{OC_{1}}$ and $\overline{OD_{1}}$. In combination with the
numerical simulation result of the radial dependence of SEP
intensity-time profiles and SEP peak intensities
($I_{max}(r)=kr^{-\alpha}$) as shown in Figure \ref{Imax-radial} and
discussed above, we can readily infer that the SEP intensity (both
entire intensity and peak intensity) observed at position $D_{1}$ is
larger than that observed at position $C_{1}$. According to the
intensity equivalence relationship between positions D and $D_{1}$
as well as that between positions C and $C_{1}$, we can infer that
the SEP intensity (both entire intensity and peak intensity)
observed by the western spacecraft D is larger than that observed by
the eastern spacecraft C. As a natural result of interplay between
perpendicular diffusion and parallel diffusion, the observation
location of the highest intensity-time profile and the largest peak
intensity of SEPs will shift from the place of best magnetic
connection and slightly toward the west of the associated SEP
source. We note that these results are valid for arbitrary energies
of energetic particles and for arbitrary particle species including
electrons and heavy ions.

\begin{figure}
 \centering
 \includegraphics[width=\columnwidth]{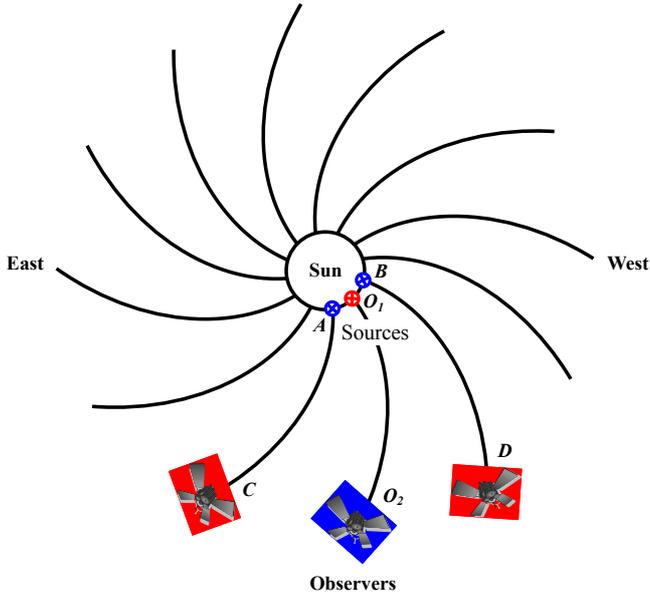}
 \caption{Diagram to illustrate the relationship between the two physical
scenarios of ``multiple SEP events with one spacecraft" (sources A
and B and spacecraft $O_{2}$ labeled in blue) and ``one SEP event
with multiple spacecraft" (source $O_{1}$ and spacecraft C and D
labeled in red). The relative position between source A (B) and
spacecraft $O_{2}$ is equivalent to the relative position between
source $O_{1}$ and spacecraft D (C). With this ``spiral
transformation trick", the two physical scenarios are unified and
identified with each other.}
      \label{Spiral-transformation}
\end{figure}

Figure \ref{Spiral-transformation} presents the relationship between
our physical scenario and that of \citet{Dresing2014}. The SEP
sources A and B and spacecraft $O_{2}$ labeled in blue indicate the
``multiple SEP events with one spacecraft" scenario, and the SEP
source $O_{1}$ and spacecraft C and D labeled in red indicate the
``one SEP event with multiple spacecraft" scenario. We emphasize the
importance of the concept of ``relative position" in understanding
the physical relationship between our results and those of
\citet{Dresing2014}. As shown in Figure \ref{Spiral-transformation},
the relative position between SEP source A and spacecraft $O_{2}$ is
equivalent to the relative position between source $O_{1}$ and
spacecraft D; the relative position between source B and spacecraft
$O_{2}$ is equivalent to the relative position between source
$O_{1}$ and spacecraft C. Therefore, the intensity observed by
spacecraft $O_{2}$ of SEPs originating from source A is equivalent
to the intensity observed by spacecraft D of SEPs originating from
source $O_{1}$; the intensity observed by spacecraft $O_{2}$ of SEPs
originating from source B is equivalent to the intensity observed by
spacecraft C of SEPs originating from source $O_{1}$. Accordingly,
with the same longitudinal separation between the solar source and
the magnetic footpoint of the observer, the following two results
are essentially the same: (1) in the scenario of ``multiple SEP
events with one spacecraft", the SEP intensity originating from the
solar source (e.g., source A in Figure \ref{Spiral-transformation})
located at the eastern side of observer footpoint is larger than the
SEP intensity originating from the source (e.g., source B) located
at the western side; (2) in the scenario of ``one SEP event with
multiple spacecraft", the SEP intensity observed by the spacecraft
(e.g., spacecraft D) located at the western side of the SEP source
is larger than the SEP intensity detected by the spacecraft (e.g.,
spacecraft C) located at the eastern side. By performing such
``spiral transformation trick" as shown in Figure
\ref{Spiral-transformation}, the two physical scenarios presented in
our works \citep{He2011,He2013,He2015a} and in \citet{Dresing2014}
are essentially unified and identified with each other.

\begin{figure}
 \centering
 \includegraphics[width=\columnwidth]{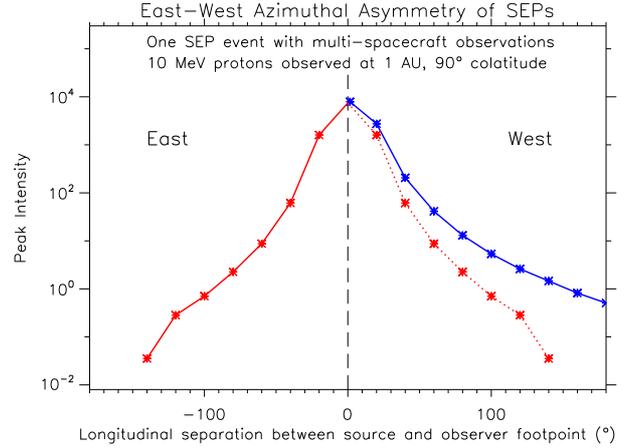}
 \caption{Multidimensional numerical modeling of the East-West longitudinal
(azimuthal) asymmetry of SEP distribution in the scenario of ``one
SEP event with multiple spacecraft". The red (blue) solid line on
the left (right) indicates the peak intensities of $10$ MeV protons
observed by spacecraft located at the eastern (western) side of the
solar source. The red dotted line on the right is mirrored from the
left. It can be clearly seen that the SEP peak intensities observed
by spacecraft located at the western side of the SEP source are
systematically larger than those observed by spacecraft located at
the eastern side. In addition, the location of the largest SEP peak
intensity slightly shifts to the west of the SEP source.}
      \label{ew-10MeV}
\end{figure}

As an example, Figure \ref{ew-10MeV} presents the multidimensional
numerical modeling result of the East-West longitudinal (azimuthal)
asymmetry of SEPs in the scenario of ``one SEP event with multiple
spacecraft". For more details on the numerical model and numerical
method of the five-dimensional Fokker-Planck focused transport
equation, we refer the reader to our simulation works
\citep[e.g.,][]{He2011,He2015a,He2015b}. In this work, we only
present Figure \ref{ew-10MeV} to demonstrate the relationship
between our results and those of \citet{Dresing2014}. The red solid
line on the left-hand half in Figure \ref{ew-10MeV} indicates the
peak intensities of $10$ MeV protons observed by a series of
spacecraft located at the eastern side of the nominal magnetic field
line connecting the solar source. On the right-hand half of Figure
\ref{ew-10MeV}, the blue solid line indicates the peak intensities
of $10$ MeV protons detected by a series of spacecraft located at
the western side of the nominal field line connecting the SEP
source. For direct comparison between the two halves, we further
mirror the SEP peak intensities from the left-hand half to the
right-hand half with a red dotted line. After this operation, we can
clearly see that the SEP peak intensities observed by spacecraft
located at the western side of the SEP source are systematically
larger than those observed by spacecraft located at the eastern
side. According to the discussions above and our numerical
simulations, actually, during the entire evolution process, the SEP
intensity observed at the western side is larger than that observed
at the eastern side, provided that the magnetic footpoints of the
pair of observers located at the western and eastern sides,
respectively, are with the same longitudinal separation relative to
the SEP source. In addition, due to the interplay between
perpendicular diffusion and parallel diffusion, the longitudinal
location at which the largest SEP peak intensity is observed
slightly shifts to somewhere on the western side of the SEP source.
Basically, the accurate extent of west-shift depends on the ratio of
perpendicular to parallel diffusion coefficients, geometry of the
Parker magnetic spiral (corresponding to solar wind speed), particle
energy and species, coverage of the SEP sources, and particle
spatial distribution in the sources.

\section{Summary and Conclusions}

\citet{He2011} reported that with the same longitudinal separation
between magnetic footpoint of observer and SEP sources, the SEPs
produced from the sources located east are detected earlier with
higher fluxes than those associated with the sources located west
\citep[see also][]{He2013,He2015a}. Our statistical survey of solar
proton events showed that with the same longitudinal separation
between the observer's magnetic footpoint and the SEP sources, the
number of SEP events associated with the sources located on the
eastern side of the observer's magnetic footpoint is larger than the
number of the SEP events originating from the sources located on the
western side. Generally, the probability of a solar proton event
being observed by spacecraft in the interplanetary space depends on
the SEP intensity in the event as well as the instrumental
sensitivity threshold of the spacecraft. Therefore, our statistical
analysis of SEP events accumulated over 16 years and our numerical
simulations of the five-dimensional Fokker-Planck transport equation
modeling taking into account perpendicular diffusion confirm each
other. We further point out that the statistical results of solar
proton events accumulated over a long time period provide an
observational evidence for the effects of perpendicular diffusion on
the SEP propagation in the interplanetary space.

We discuss in detail the effects of perpendicular diffusion on the
formation of the East-West longitudinally (azimuthally) asymmetric
distribution of SEPs. The role of perpendicular diffusion in the
physical scenarios of both ``multiple SEP events with one
spacecraft" and ``one SEP event with multiple spacecraft" is
thoroughly demonstrated. We obtain the radial dependence of SEP peak
intensities $I_{max}(r)=kr^{-1.7}$ from numerically modeling the
five-dimensional Fokker-Planck focused transport equation. We employ
this functional relation ($I_{max}(r)=kr^{-1.7}$) to quantitatively
illustrate and explain the East-West longitudinal (azimuthal)
asymmetry of SEP distribution in the heliosphere. The inherent
relationship between our results \citep{He2011,He2013,He2015a} and
those of \citet{Dresing2014} is comprehensively discussed. In the
context of perpendicular diffusion, our results and those of
\citet{Dresing2014} are essentially unified and identified with each
other. We note that the SEP phenomenon of longitudinally
(azimuthally) asymmetric distribution (``East-West azimuthal
asymmetry") cannot be decently explained by any other hypothetical
mechanisms including the so-called reflecting boundary (magnetic
mirror) and diffusion barrier (solar wind structure).

In this work, we primarily discuss the SEP scenario in the ecliptic
plane. It should be noted that the results and the conclusions
presented above essentially hold for arbitrary heliospheric
latitudes. As a special and unique SEP phenomenon, the East-West
longitudinally (azimuthally) asymmetric distribution of SEPs
represents the important effects of the perpendicular diffusion
processes (which was usually ignored in previous studies, especially
in the observational community) and of the large-scale Parker spiral
geometry of the interplanetary magnetic field. In addition, we note
that the East-West longitudinal (azimuthal) asymmetry of SEP
distribution is a universal phenomenon independent of particle
energy, particle species, particle origin (on or near the Sun or in
the interplanetary space), and observer's location (latitude,
longitude, and radial distance) in the heliosphere. Therefore, our
results presented in this work will be valuable for understanding
the Sun-Earth relations and especially for understanding the
three-dimensional propagation of SEPs in the heliosphere.

\section*{Acknowledgements}

This work was supported in part by the National Natural Science
Foundation of China under grants 41204130, 41474154, 41321003, and
41131066, the National Important Basic Research Project under grant
2011CB811405, the Chinese Academy of Sciences under grant
KZZD-EW-01-2, and the Open Research Program from Key Laboratory of
Geospace Environment, Chinese Academy of Sciences. H.-Q. He
gratefully acknowledges the partial support of the International
Postdoctoral Exchange Fellowship Program of China under grant
20130023, and the K. C. Wong Education Foundation. We benefited from
the solar proton events data provided by the Space Weather
Prediction Center of NOAA and the instrument teams of GOES
spacecraft. We thank the NASA/Space Physics Data Facility
(SPDF)/CDAWeb for providing the solar wind data of SOHO and ACE.





\begin{thebibliography}{99}
\bibitem[Beeck \& Wibberenz(1986)]{Beeck1986}Beeck, J., \& Wibberenz, G. 1986, \apj, 311, 437
\bibitem[Beeck et al.(1987)]{Beeck1987}Beeck, J., Mason, G. M., Hamilton, D. C., et al. 1987, \apj, 322, 1052
\bibitem[Cliver \& Cane(2002)]{Cliver2002}Cliver, E. W., \& Cane, H. V. 2002, EOS Trans. AGU, 83, 61
\bibitem[Dresing et al.(2014)]{Dresing2014}Dresing, N., G\'{o}mez-Herrero, R., Heber, B., et al. 2014, \aap, 567, A27
\bibitem[Dresing et al.(2012)]{Dresing2012}Dresing, N., G\'{o}mez-Herrero, R., Klassen, A., Heber, B., Kartavykh, Y., \& Dr\"oge, W. 2012, \solphys, 281, 281
\bibitem[Dr\"oge et al.(2014)]{Droge2014}Dr\"oge, W., Kartavykh, Y. Y., Dresing, N., Heber, B., \& Klassen, A. 2014, \jgr, 119, 6074
\bibitem[Dr\"oge et al.(2010)]{Droge2010}Dr\"oge, W., Kartavykh, Y. Y., Klecker, B., \& Kovaltsov, G. A. 2010, \apj, 709, 912
\bibitem[Effenberger et al.(2012)]{Effenberger2012}Effenberger, F., Fichtner, H., Scherer, K., et al. 2012, \apj, 750, 108
\bibitem[Giacalone \& Jokipii(2012)]{Giacalone2012}Giacalone, J., \& Jokipii, J. R. 2012, \apjl, 751, L33
\bibitem[Hasselmann \& Wibberenz(1968)]{Hasselmann1968}Hasselmann, K., \& Wibberenz, G. 1968, Z. Geophys., 34, 353
\bibitem[Hasselmann \& Wibberenz(1970)]{Hasselmann1970}Hasselmann, K., \& Wibberenz, G. 1970, \apj, 162, 1049
\bibitem[He et al.(2011)]{He2011}He, H.-Q., Qin, G., \& Zhang, M. 2011, \apj, 734, 74
\bibitem[He \& Wan(2013)]{He2013}He, H.-Q., \& Wan, W. 2013, Proc. ICRC (Brazil), paper 0267, arXiv:1502.03090
\bibitem[He \& Wan(2015)]{He2015a}He, H.-Q., \& Wan, W. 2015, \apjs, 218, 17
\bibitem[He(2015)]{He2015b}He, H.-Q. 2015, \apj, 814, 157
\bibitem[Jokipii(1966)]{Jokipii1966}Jokipii, J. R. 1966, \apj, 146, 480
\bibitem[Kelly et al.(2012)]{Kelly2012}Kelly, J., Dalla, S., \& Laitinen, T. 2012, \apj, 750, 47
\bibitem[Lario et al.(2013)]{Lario2013}Lario, D., Aran, A., G\'{o}mez-Herrero, R., et al. 2013, \apj, 767, 41
\bibitem[Parker(1958)]{Parker1958}Parker, E. N. 1958, \apj, 128, 664
\bibitem[Roelof et al.(1992)]{Roelof1992}Roelof, E. C., Gold, R. E., Simnett, G. M., Tappin, S. J., Armstrong, T. P., \& Lanzerotti, L. J. 1992, \grl, 19, 1243
\bibitem[Shalchi(2005)]{Shalchi2005}Shalchi, A. 2005, \mnras, 363, 107
\bibitem[Strauss \& Fichtner(2015)]{Strauss2015}Strauss, R. D., \& Fichtner, H. 2015, \apj, 801, 29
\bibitem[Veselovsky et al.(2004)]{Veselovsky2004}Veselovsky, I. S., Panasyuk, M. I., Avdyushin, S. I., et al. 2004, Cosmic Research, 42, 435
\bibitem[Zhang et al.(2009)]{Zhang2009}Zhang, M., Qin, G., \& Rassoul, H. 2009, \apj, 692, 109
\bibitem[Zhao et al.(2007)]{Zhao2007}Zhao, X., Feng, X., \& Wu, C.-C. 2007, \jgr, 112, A06107
\end{thebibliography}




\appendix

\section{List of Solar Proton Events}

We list here the solar proton events analyzed in this study. See
Table \ref{table:events} for details.

\clearpage

\begin{table}
\centering

\caption{Solar Proton Events Affecting the Earth Environment
(1996-2011)(collected from Space Weather Prediction Center of NOAA)
\label{table:events}}

\newsavebox{\tablebox}
\begin{lrbox}{\tablebox}

\begin{tabular}{|ccc|c|cccc|ccc|}

\hline

\multicolumn{3}{|c|}{Solar Proton Event}&Associated CME&\multicolumn{4}{|c|}{Flare and Active Region}&\multicolumn{3}{|c|}{Solar Wind Speed}\\
Start&Maximum&Proton Flux&Onset&Flare Maximum&Importance&Location&Region No.&SOHO&ACE&Mean\\
(Day/UT)&(Day/UT)&(pfu~@~$>10$~MeV)&(Direction/Day/UT)&(Day/UT)&(X Ray/Opt.)&&&\multicolumn{3}{|c|}{$(km~s^{-1})$}\\

\hline

1997.11.04/08:30&1997.11.04/11:20&72&W/04/06:10&1997.11.04/05:58&X2/2B&S14W33&8100&343&&\\
1997.11.06/13:05&1997.11.07/02:55&490&W/06/$>$13:00&1997.11.06/11:55&X9/2B&S18W63&8100&425&&\\
1998.04.20/14:00&1998.04.21/12:05&1700&W/20/10:07&1998.04.20/10:21&M1/EPL&S43W90&8194&377&354&366\\
1998.05.02/14:20&1998.05.02/16:50&150&Halo/02/14:06&1998.05.02/13:42&X1/3B&S15W15&8210&569&557&563\\
1998.05.06/08:45&1998.05.06/09:45&210&W/06/08:29&1998.05.06/08:09&X2/1N&S11W65&8210&437&461&449\\
1998.08.24/23:55&1998.08.26/10:55&670&NA&1998.08.24/22:12&X1/3B&N30E07&8307&&651&\\
1998.09.25/00:10&1998.09.25/01:30&44&NA&1998.09.23/07:13&M7/3B&N18E09&8340&&713&\\
1998.09.30/15:20&1998.10.01/00:25&1200&NA&1998.09.30/13:50&M2/2N&N23W81&8340&&456&\\
1998.11.14/08:10&1998.11.14/12:40&310&NA&1998.11.14/05:18&C1/BSL&N28W90&8375?&&407&\\
1999.01.23/11:05&1999.01.23/11:35&14&NA&1999.01.20/20:04&M5&N27E90&&&588&\\
1999.05.05/18:20&1999.05.05/19:55&14&Halo/03/06:06&1999.05.03/06:02&M4/2N&N15E32&8525&448&423&436\\
1999.06.04/09:25&1999.06.04/10:55&64&NW/04/07:26&1999.06.04/07:03&M3/2B&N17W69&8552&416&408&412\\
2000.02.18/11:30&2000.02.18/12:15&13&W/18/09:54&2000.02.17/20:35&M1/2N&S29E07&8872&368&&\\
2000.04.04/20:55&2000.04.05/09:30&55&W/04/16:32&2000.04.04/15:41&C9/2F&N16W66&8933&394&382&388\\
2000.06.07/13:35&2000.06.08/09:40&84&Halo/06/15:54&2000.06.06/15:25&X2/3B&N20E18&9026&689&704&697\\
2000.06.10/18:05&2000.06.10/20:45&46&Halo/10/17:08&2000.06.10/17:02&M5/3B&N22W38&9026&479&468&474\\
2000.07.14/10:45&2000.07.15/12:30&24000&Halo/14/10:54&2000.07.14/10:24&X5/3B&N22W07&9077&398&&\\
2000.07.22/13:20&2000.07.22/14:05&17&NW/22/12:30&2000.07.22/11:34&M3/2N&N14W56&9085&412&411&412\\
2000.09.12/15:55&2000.09.13/03:40&320&Halo/12/13:31&2000.09.12/12:13&M1/2N&S17W09&Filament&397&416&407\\
2000.10.16/11:25&2000.10.16/18:40&15&Halo/16/07:27&2000.10.16/07:28&M2&N04W90&9182?&527&544&536\\
2000.10.26/00:40&2000.10.26/03:40&15&Halo/25/08:26&2000.10.25/11:25&C4&N00W90&&395&390&393\\
2000.11.24/15:20&2000.11.26/20:30&940&Halo/24/05:30&2000.11.24/05:02&X2/3B&N20W05&9236&530&585&558\\
2001.01.28/20:25&2001.01.29/06:55&49&Halo/28/15:54&2001.01.28/16:00&M1/1N&S04W59&9313&392&395&394\\
2001.03.29/16:35&2001.03.30/06:10&35&Halo/29/10:26&2001.03.29/10:15&X1/1N&N14W12&9393&&450&\\
2001.04.02/23:40&2001.04.03/07:45&1110&NW/02/~22:00&2001.04.02/21:51&X20&N18W82&9393&443&480&462\\
2001.04.10/08:50&2001.04.11/20:55&355&Halo/10/05:30&2001.04.10/05:26&X2/3B&S23W09&9415&&724&\\
2001.04.15/14:10&2001.04.15/19:20&951&W/15/14:30&2001.04.15/13:50&X14/2B&S20W85&9415&409&496&453\\
2001.04.28/04:30&2001.04.28/05:00&57&Halo/26/12:30&2001.04.26/13:12&M7/2B&N17W31&9433&644&619&632\\
2001.09.15/14:35&2001.09.15/14:55&11&SW/15/11:54&2001.09.15/11:28&M1/1N&S21W49&9608&518&586&552\\
2001.09.24/12:15&2001.09.25/22:35&12900&Halo/24/10:30&2001.09.24/10:38&X2/2B&S16E23&9632&418&&\\
2001.10.01/11:45&2001.10.02/08:10&2360&SW/01/05:30&2001.10.01/05:15&M9&S22W91&9628&445&483&464\\
2001.10.19/22:25&2001.10.19/22:35&11&Halo/19/16:50&2001.10.19/16:30&X1/2B&N15W29&9661&352&344&348\\
2001.10.22/19:10&2001.10.22/21:30&24&SE/22/18:26&2001.10.22/17:59&X1/2B&S18E16&9672&532&528&530\\
2001.11.04/17:05&2001.11.06/02:15&31700&Halo/04/16:35&2001.11.04/16:20&X1/3B&N06W18&9684&399&413&406\\
2001.11.19/12:30&2001.11.20/00:10&34&Halo/17/05:30&2001.11.17/05:25&M2/1N&S13E42&9704&522&509&516\\
2001.11.22/23:20&2001.11.24/05:55&18900&Halo/22/23:30&2001.11.22/23:30&M9/2N&S15W34&9704&428&467&448\\
2001.12.26/06:05&2001.12.26/11:15&779&W/26/05:30&2001.12.26/05:40&M7/1B&N08W54&9742&392&383&388\\
2001.12.29/05:10&2001.12.29/08:15&76&E/28/20:06&2001.12.28/20:45&X3&S26E90&9767&433&439&436\\
2002.02.20/07:30&2002.02.20/07:55&13&W/20/06:30&2002.02.20/06:12&M5/1N&N12W72&9825&&400&\\
2002.03.17/08:20&2002.03.17/08:50&13&Halo/15/23:06&2002.03.15/23:10&M2/1F&S08W03&9866&&278&\\

\hline

\end{tabular}

\end{lrbox}
\resizebox{1.0\textwidth}{!}{\usebox{\tablebox}}

\end{table}

\clearpage


\begin{table}
\centering

\begin{lrbox}{\tablebox}

\begin{tabular}{|ccc|c|cccc|ccc|}

\hline

\multicolumn{3}{|c|}{Solar Proton Event}&Associated CME&\multicolumn{4}{|c|}{Flare and Active Region}&\multicolumn{3}{|c|}{Solar Wind Speed}\\
Start&Maximum&Proton Flux&Onset&Flare Maximum&Importance&Location&Region No.&SOHO&ACE&Mean\\
(Day/UT)&(Day/UT)&(pfu~@~$>10$~MeV)&(Direction/Day/UT)&(Day/UT)&(X Ray/Opt.)&&&\multicolumn{3}{|c|}{$(km~s^{-1})$}\\

\hline

2002.03.20/15:10&2002.03.20/15:25&19&W/18/02:54&2002.03.18/02:31&M1&S09W46&~9866&&555&\\
2002.04.17/15:30&2002.04.17/15:40&24&Halo/17/08:26&2002.04.17/08:24&M2/2N&S14W34&9906&&520&\\
2002.04.21/02:25&2002.04.21/23:20&2520&W/21/01:27&2002.04.21/01:51&X1/1F&S14W84&9906&&454&\\
2002.05.22/17:55&2002.05.23/10:55&820&Halo/22/03:26&2002.05.22/03:54&C5/DSF&S19W56&&&613&\\
2002.07.16/17:50&2002.07.17/16:00&234&Halo/15/20:30&2002.07.15/20:08&X3/3B&N19W01&30&&476&\\
2002.08.14/09:00&2002.08.14/16:20&26&NW/14/02:06&2002.08.14/02:12&M2/1N&N09W54&61&&489&\\
2002.08.22/04:40&2002.08.22/09:40&36&SW/22/02:00&2002.08.22/01:57&M5/2B&S07W62&69&&439&\\
2002.08.24/01:40&2002.08.24/08:35&317&W/24/01:27&2002.08.24/01:12&X3/1F&S08W90&69&&375&\\
2002.09.07/04:40&2002.09.07/16:50&208&Halo/05/16:54&2002.09.05/17:06&C5/DSF&N09E28&102&&500&\\
2002.11.09/19:20&2002.11.10/05:40&404&SW/09/13:31&2002.11.09/13:23&M4/2B&S12W29&180&&376&\\
2003.05.28/23:35&2003.05.29/15:30&121&Halo/28/00:50&2003.05.28/00:27&X3/2B&S07W17&365&&653&\\
2003.05.31/04:40&2003.05.31/06:45&27&W/31/02:30&2003.05.31/02:24&M9/2B&S07W65&365&&752&\\
2003.06.18/20:50&2003.06.19/04:50&24&Halo/17/23:30&2003.06.17/22:55&M6&S08E61&386&&585&\\
2003.10.26/18:25&2003.10.26/22:35&466&Halo/26/17:54&2003.10.26/18:19&X1/1N&N02W38&484&&482&\\
2003.10.28/12:15&2003.10.29/06:15&29500&Halo/28/10:54&2003.10.28/11:10&X17/4B&S16E08&486&&790&\\
2003.11.04/22:25&2003.11.05/06:00&353&Halo/04/1954&2003.11.04/19:29&X28/3B&S19W83&486&&580&\\
2003.11.21/23:55&2003.11.22/02:30&13&SW/21/00:26&2003.11.20/23:53&M5/2B&N02W17&501&&494&\\
2004.04.11/11:35&2004.04.11/18:45&35&SW/11/04:30&2004.04.11/04:19&C9/1F&S14W47&588&&440&\\
2004.07.25/18:55&2004.07.26/22:50&2086&Halo/25/15:30&2004.07.25/15:14&M1/1F&N08W33&652&&783&\\
2004.09.13/21:05&2004.09.14/00:05&273&Halo/12/00:36&2004.09.12/00:56&M4/2N&N04E42&672&&554&\\
2004.09.19/19:25&2004.09.20/01:00&57&W/19/22:24&2004.09.19/17:12&M1&N03W58&672&&387&\\
2004.11.07/19:10&2004.11.08/01:15&495&Halo/07/17:06&2004.11.07/16:06&X2&N09W17&696&&642&\\
2005.01.16/02:10&2005.01.17/17:50&5040&Halo/15/23:06&2005.01.15/23:02&X2&N15W05&720&&586&\\
2005.05.14/05:25&2005.05.15/02:40&3140&Halo/13/17:22&2005.05.13/16:57&M8/2B&N12E11&759&&556&\\
2005.06.16/22:00&2005.06.17/05:00&44&W/16/20:03&2005.06.16/20:22&M4&N09W87&775&&592&\\
2005.07.14/02:45&2005.07.15/03:45&134&Halo/13/14:30&2005.07.13/14:49&M5&N10W80&786&&430&\\
2005.07.27/23:00&2005.07.29/17:15&41&Halo/27/04:54&2005.07.27/05:02&M3&N11E90&792&&554&\\
2005.08.22/20:40&2005.08.23/10:45&330&Halo/22/17:30&2005.08.22/17:27&M5/1N&S12W60&798&&457&\\
2005.09.08/02:15&2005.09.11/04:25&1880&E/07/17:23&2005.09.07/17:40&X17/3B&S06E89&808&&873&\\
2006.12.06/15:55&2006.12.07/19:30&1980&Halo&2006.12.05/10:35&X9/2N&S07E79&930&&580&\\
2006.12.13/03:10&2006.12.13/09:25&698&Halo/13/02:54&2006.12.13/02:40&X3/4B&S05W23&930&&641&\\
2010.08.14/12:30&2010.08.14/12:45&14&W/14/13:25&2010.08.14/10:05&C4/0F&N17W52&1099&&409&\\
2011.03.08/01:05&2011.03.08/08:00&50&NW/07/20:00&2011.03.07/20:12&M3/SF&N24W59&1164&&379&\\
2011.06.07/08:20&2011.06.07/18:20&72&SW/08/17:50&2011.06.07/08:03&M2/2N&S21W64&1226&&457&\\
2011.08.04/06:35&2011.08.05/21:50&96&NW/06/05:15&2011.08.04/04:12&M9/2B&N15W49&1261&&583&\\
2011.08.09/08:45&2011.08.09/12:10&26&NW/09/17:10&2011.08.09/08:05&X6/2B&N17W83&1263&&588&\\
2011.09.23/22:55&2011.09.26/11:55&35&NE/27/04:30&2011.09.22/11:01&X1/2N&N11E74&1302&&428&\\
2011.11.26/11:25&2011.11.27/01:25&80&NW/28/01:45&2011.11.26/07:10&C1&N08W49&1353&&358&\\

\hline

\end{tabular}

\end{lrbox}
\resizebox{1.0\textwidth}{!}{\usebox{\tablebox}}

\end{table}

\clearpage


\bsp    
\label{lastpage}
\end{document}